\documentclass[final
]{dmtcs}
\usepackage[round]{natbib}

\usepackage{bbm,amsmath,amssymb}
\newtheorem{theorem}{Theorem}[section]
\newtheorem{lemma}{Lemma}[section]
\newtheorem{proposition}{Proposition}[section]

\newcommand{\ZZ}{{\mathbbm{Z}}}
\newcommand{\NN}{{\mathbbm{N}}}

\newcommand{\B}{{\mathcal B}}
\newcommand{\G}{{\mathcal G}}
\newcommand{\calS}{{\mathcal S}}
\newcommand{\f}{{\mathbf f}}

\newcommand{\card}{\mathop{\mathrm{card}}}

\author{Henryk Fuk\'s \thanks{The author acknowledges partial financial support from Natural Sciences and Engineering Research Counclil of Canada, in the form of a Discovery Grant.}}
\title[Cellular automaton rule 172]{Probabilistic initial value problem for cellular automaton rule 172}
\address{Department of Mathematics, Brock University, St. Catharines, ON L2S 3A1, Canada}
\received{5 Apr 2010}
\begin{document}
\maketitle
\begin{abstract}
We present a method of solving of the probabilistic initial value problem for cellular automata (CA) using  CA rule 172 as an example. For a disordered
initial condition on an infinite lattice, we derive exact expressions for the density of ones at arbitrary time step. In order to do this,
we analyze topological structure of preimage trees of finite strings of length 3. Level sets of these trees can
be enumerated directly using classical combinatorial methods, yielding expressions for the number of $n$-step preimages
of all strings of length 3, and, subsequently, probabilities of occurrence of these strings in a configuration obtained
from the initial one after $n$ iterations of rule 172. The density of ones can be expressed in terms of Fibonacci numbers, while
expressions for probabilities of other strings involve Lucas numbers. Applicability of this method to other CA rules is briefly discussed.
  \end{abstract}

\section{Introduction}
While working on a certain problem in \textit{complexity engineering}, that is, trying to construct a cellular automaton rule  performing
some useful computational task, the author encountered the following question.
Let   $f:\{0,1\}^3 \to \{0,1\}$ be defined as
 \begin{equation}\label{selector}
 f(x_1,x_2,x_3) = \left \{ \begin{array}{ll}
                              x_2     & \mbox{if $x_1=0$,}\\
                              x_3     & \mbox{if $x_1=1$.}
                    \end{array}
 \right.
\end{equation}
This function may be called \emph{selective copier}, since it returns (copies) one of its inputs $x_2$ or $x_3$ depending on the state of the first input variable $x_1$. Suppose now that $s$ be a bi-infinite sequence of binary symbols, i.e., $s=\ldots s_{-2}s_{-1}s_0s_1s_2\ldots$, $i\in \ZZ$. We will transform this string using
the selective copier, that is, for each $i$, we keep  $s_i$ if it is preceded by $0$, or replace it by
$s_{i+1}$ otherwise, so that each $s_i$ is simultaneously  replaced by $f(s_{i-1},s_i,s_{i+1})$. Consider now the question:
\textit{Assuming that the initial sequence is randomly generated, what is 
the proportion of 1's in the sequence after $n$ iterations of the aforementioned procedure?}

Function defined by eq. ($\ref{selector}$) is a local function of cellular automaton rule 172, using Wolfram numbering, and the aforementioned question
is an example
of a broader class of problems, which could be called probabilistic initial value problems for cellular automata: given initial
distribution of \textit{infinite configurations}, what is the probability of occurrence of a given \textit{finite string} in a configuration obtained from the
initial one by $n$ iterations of the cellular automaton rule? In what follows, we will demonstrate how one can approach probabilistic initial value problem using cellular automaton rule 172 as an example. 

\section{Basic definitions}
Let $\G=\{0,1,...N-1\}$ be called {\em a symbol set}, and let $\calS(\G)$
 be the set of all bisequences over $\G$, where by a bisequence we mean a
 function on  $\ZZ$ to $\G$. Set  $\calS(\G)$ will be
called {\em the configuration space}. Throughout the remainder of this
text
 we shall  assume that $\G=\{0,1\}$, and  the configuration space
 $\calS(\G)=\{0,1\}^{\ZZ}$ will be simply denoted by $\calS$.

{\em A block of length} $n$ is an ordered set $b_{0} b_{1}
\ldots b_{n-1}$, where $n\in \NN$, $b_i \in \G$.
Let $n\in \NN$ and let
$\B_n$ denote the set of all blocks of length $n$ over $\G$ and $\B$ be
the set of all finite blocks over $\G$.

For $r \in \NN$, a mapping $f:\{0,1\}^{2r+1}\mapsto\{0,1\}$ will be called {\em a cellular
 automaton rule of radius $r$}. Alternatively, the function $f$ can be
 considered as a mapping of $\B_{2r+1}$ into $\B_0=\G=\{0,1\}$. 

Corresponding to $f$ (also called {\em a local mapping}) we define a
 {\em global mapping}  $F:\calS \to \calS$ such that
$
(F(s))_i=f(s_{i-r},\ldots,s_i,\ldots,s_{i+r})
$
 for any $s\in \calS$.

A {\em block evolution operator} corresponding to $f$ is a mapping
 $\f:\B \mapsto \B$ defined as follows. 
Let $r\in \NN$ be the radius of $f$, and let  $a=a_0a_1 \ldots a_{n-1}\in \B_{n}$
where $n \geq 2r+1 >0$. Then 
\begin{equation}
\f(a) = \{ f(a_i,a_{i+1},\ldots,a_{i+2r})\}_{i=0}^{n-2r-1}.
\end{equation}
 Note that if
$b \in B_{2r+1}$ then $f(b)=\f(b)$.

We will consider the case of $\G=\{0,1\}$ and $r=1$ rules,
 i.e., {\em elementary cellular automata}. In this case, when $b\in\B_3$,
then $f(b)=\f(b)$. The set 
$\B_3=\{000,001,010,011,100,101,101,110$, $111\}$ will be called the set of \textit{basic blocks}.

The number of $n$-step preimages of the block $b$ under the rule $f$
is defined as the number of elements of the set $\f^{-n}(b)$.
Given an elementary rule $f$, we will be especially interested in
the number of $n$-step preimages of basic blocks
under the rule $f$.

\section{Probabilistic initial value problem}
The appropriate mathematical description of an initial distribution of
configurations is a probability measure $\mu$ on $\calS$. 
Such a measure can be formally constructed as follows. 
If $b$ is a block of length $k$, i.e.,
$b=b_0b_1 \ldots b_{k-1}$, then for $i \in \ZZ$ we define a cylinder set.
The cylinder set is a set of all possible configurations with fixed
values at a finite number of sites.  Intuitively, measure of the
cylinder  set given by the block $b=b_0\ldots b_{k-1}$, denoted by
 $\mu[C_i(b)]$, is simply a probability of occurrence
of the block $b$ in a place starting at $i$. If the measure $\mu$ is shift-invariant, than
$\mu(C_i(b))$ is independent of $i$, and we will therefore drop the
index $i$ and  simply write  $\mu(C(b))$.

The Kolmogorov consistency theorem
states that every probability measure $\mu$ satisfying the consistency
condition
\begin{equation}\label{consist}
\mu [C_i(b_1 \ldots b_k)]= \mu [C_i(b_1 \ldots b_k,0)]+\mu [C_i(b_1 \ldots  b_k,1)]
\end{equation}
extends to a shift invariant measure on $\cal S$ \citep{Dynkin69} .For $p\in[0,1]$, the Bernoulli
measure defined as $\mu_p[C(b)]=p^j(1-p)^{k-j}$, where $j$ is a number
of ones in $b$ and $k-j$ is a number of zeros in $b$, is an example of
such a shift-invariant (or spatially homogeneous) measure. It describes a set of random
configurations with the probability $p$ that a given site is in state $1$.

Since a cellular automaton rule with global function $F$ maps a configuration in $\calS$
to another configuration in $\calS$, we can define the action of $F$ on measures
on $\calS$.
For all measurable subsets $E$ of $\calS$ we define
$(F\mu)(E)=\mu(F^{-1}(E))$, where $F^{-1}(E)$ is an inverse image of
$E$ under $F$.

If the initial configuration was specified by $\mu_p$, what can be
said about $F^n\mu_p$ (i.e., what is the probability measure after $n$
iterations of $F$)? In particular, given a block $b$, what is the
probability of the occurrence of this block in a configuration obtained
from a random configuration after $n$ iterations of a given rule?

The general question of finding the iterrates of the Bernoulli measure under a given CA has been
extensively studied in recent years by many authors, including, among others, 
\cite{Lind84,Ferrari2000,Maas2003,Host2003,Pivato2002,Pivato2004,Maas2006} and \cite{Maas2006b}.
In this paper, we will approach the problem from somewhat different angle, using very elementary methods
and without resorting to advanced apparatus of ergodic theory and symbolic dynamics. We will consider iterates
of the Bernoulli measure by analyzing patterns in preimage sets. 
%

\begin{figure}[t]
  \begin{center}
    \includegraphics[scale=0.8]{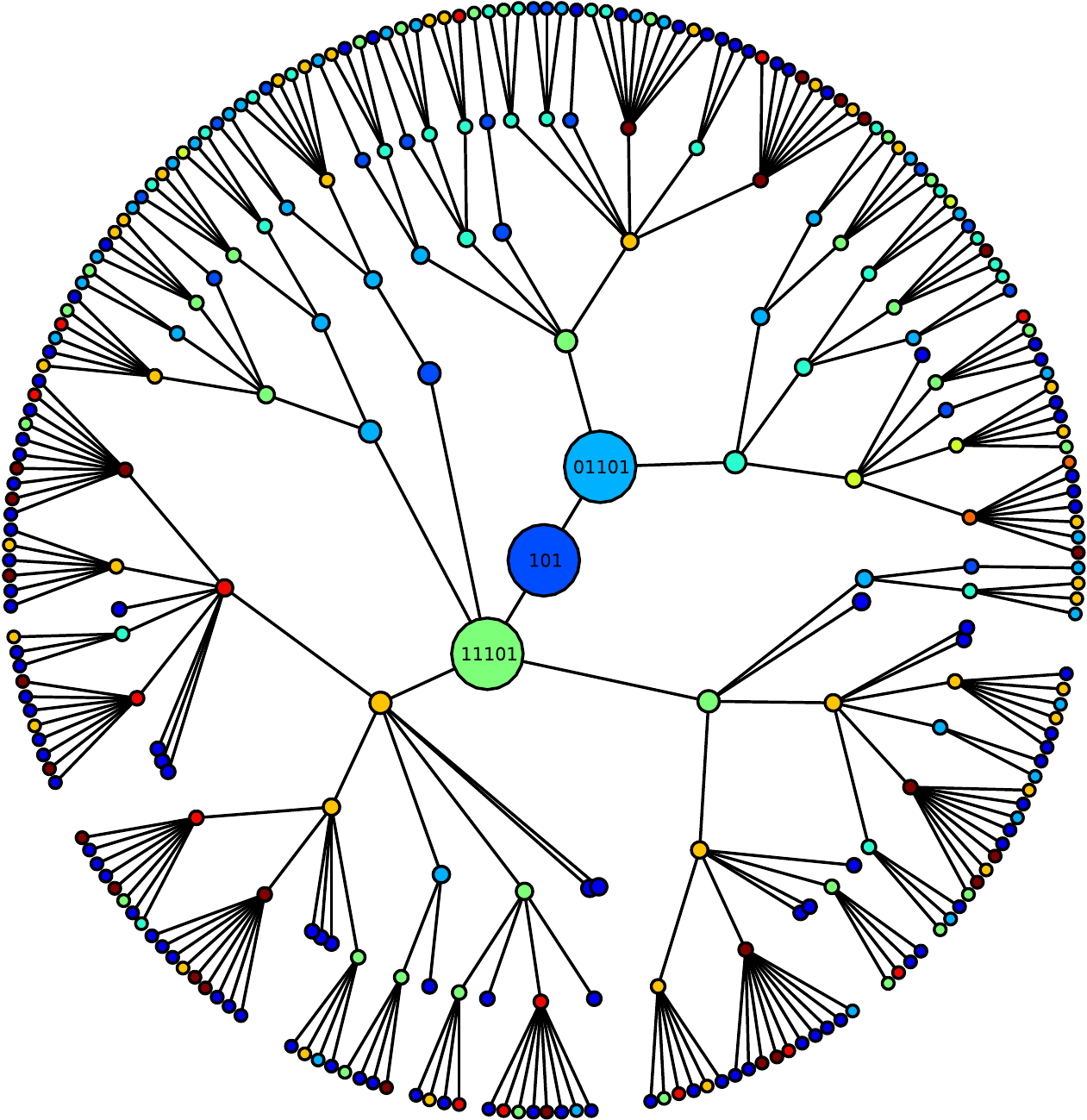}
  \end{center}
 \caption{Preimage tree for rule 172 rooted at $101$.}
  \label{preimtree101}
\end{figure}

For a given block $b$, the set of $n$-step preimages
is $\f^{-n}(b)$. Then, by the definition of the action of $F$
on the initial measure, we have
\begin{equation}
 (F^n\mu_p)(C(b)) =  \mu_p \left(F^{-n}(C(b)) \right),
\end{equation}
and consequently
\begin{equation} \label{blockmeasure}
  (F^n\mu_p)(C(b)) =\sum_{a \in \f^{-n}(b)}  \mu_p (a).
\end{equation}
Let us define the probability of occurrence of block $b$ in a configuration obtained from the initial one by
$n$ iterations of the CA rule as
\begin{equation}
 P_n(b)= (F^n\mu_p)(C(b)).
\end{equation}
Using this notation, eq. (\ref{blockmeasure}) becomes
\begin{equation}
 P_n(b)= \sum_{a \in \f^{-n}(b)}  P_0(a).
\end{equation}
If the initial measure is $\mu_{1/2}$, then all blocks of a given length are equally probable, and 
$P_0(a)=\frac{1}{2^{|a|}}$, where $|a|$ is the length of the block $a$. For elementary CA rule,
the length of $n$-step preimage of $b$ is $2n+|b|$, therefore
\begin{equation}\label{blockprob}
 P_n(b) = 2^{-|b|-2n} \card \f^{-n}(b).
\end{equation}
This equation tells us that if the initial measure is symmetric ($\mu_{1/2}$), then all we need to know 
in order to compute $P_n(b)$ is the cardinality of $ \f^{-n}(b)$. One way to think about this
is to draw a \textit{preimage tree} for~$b$. We start form $b$ as a root of the tree, and determine 
all its preimages. Then each of these preimages is connected with $b$ by an edge. They constitute
level 1 of the  preimage tree. Then, for each block of level 1, we again compute its preimages and we link
them with that block, thus obtaining level 2. Repeating this operation \textit{ad infinitum}, we obtain
a tree such as the one shown in~Figure \ref{preimtree101}. In that figure, five levels of the preimage tree
for rule 172 rooted at $101$ are shown, with only first level  labelled.

Note that $\card \f^{-n}(b)$ corresponds to the number of vertices in the $n$-th level of the preimage tree.
One thus only needs to know cardinalities of level sets in order to use eq. (\ref{blockprob}), while the exact topology
of connections between vertices of the preimage tree is unimportant.
The key problem, therefore, is to enumerate level sets. In order to answer the question posed in the introduction,
we need to compute $P_n(1)$ for rule 172, which, in turn, requires that we enumerate level sets 
of a preimage tree rooted at $1$. It turns out that for rule 172 the preimage tree rooted at 1 
is rather complicated, and that it is more convenient to consider preimage trees rooted at other blocks. In the next
section, we will show how to express $P_n(1)$ by some other block probabilities. From now on,
$\f$ will exclusively denote the block evolution operator for rule 172.
\section{Block probabilities}
Since $\f^{-1}(1)=\{010,011,101,111\}$, we have
$P_{n+1}(1)=P_{n}(010)+P_{n}(011)+P_{n}(101)+P_{n}(111)$.  Due to consistency conditions (eq.  \ref{consist}),
$P_{n}(010)+P_{n}(011)=P_n(01)$, and we obtain 
\begin{equation}
 P_{n+1}(1)=P_n(01) + P_{n}(101)+P_{n}(111).
\end{equation}
This can be transformed even further by noticing that $P_n(01)=P_n(001)+P_n(101)$, therefore
\begin{equation}
 P_{n}(1)=P_{n-1}(001) + 2P_{n-1}(101)+P_{n-1}(111).
\end{equation}
By using eq. (\ref{blockprob}) and defining $c_n=P_n(1)$ we obtain
\begin{equation}\label{densityformula}
 c_n=  \frac{\card \f^{-n+1}(001) + 2\card \f^{-n+1}(101) +\card \f^{-n+1}(111)}{2^{2n+1}}.
\end{equation}
This means that in order to compute $c_n$, we need to know cardinalities of
$n$-step preimages of 001, 101, and 111.
\section{Structure of preimage sets}
The structure of level sets of preimage trees rooted at $001$, $101$, and $111$ will be 
described in the following three propositions.
 \begin{proposition}\label{struct001}
 Block $b$ belongs to $\f^{-n}(001)$ if and only if it has the structure 
\begin{equation}
b= \underbrace{\star \star \ldots \star}_n  0 0 1 \underbrace{\star \star \ldots \star}_n,
\end{equation}
where $\star$ represents arbitrary symbol from the set $\{0,1\}$.
 \end{proposition}
Let us first observe that $\f^{-1}(001)=\{00010,00011,10010,10011\}$, which means
that $\f^{-1}(001)$ can be represented as $\star \star001\star \star$. Similarly, therefore,
 $\f^{-2}(001)$ has the structure  $\star \star \star001\star \star \star$, and by induction,
for any $n$, the structure of  $\f^{-n}(001)$ must be $\underbrace{\star \star \ldots \star}_n  0 0 1 \underbrace{\star \star \ldots \star}_n$. $\Box$
 \begin{proposition}\label{struct101}
 Block $b$ belongs to $\f^{-n}(101)$ if and only if it has the structure 
\begin{equation}
b= \underbrace{\star \star \ldots \star}_{n-1}  a_1 a_2\ldots a_n 1101,
\end{equation}
where  $a_i \in \{0,1\}$ for $i=1,\ldots, n$ and the string $ a_1 a_2\ldots a_n$ does not contain any pair of adjacent zeros, that
is. $a_ia_{i+1} \neq 00$ for all $i=1,\ldots, n-1$.
 \end{proposition}
Two observations will be crucial for the proof. First of all, $\f^{-1}(101)=\{01101,
11101\}$, thus $\f^{-1}(101)$ has the structure $\star1101$.  Furthermore, we have
$\f^{-1}(1101) = \{011101 , 101101 , 111101\}$, meaning that if $1101$ appears in a configuration, and is not
preceded by $00$, then after application of the rule 172,  $1101$ will still appear, but shifted one position to the left.
All this means that if $b$ is to be an $n$-step preimage of $101$, it must end with $1101$. After each application
of rule 172 to $b$, the block $1101$ will remain at the end as long as it is not preceded by two zeros.

Now, let us note that $\f^{-1}(00)=\{0000, 0001,1000,1001,1100\}$, which means that preimage of $00$
is either $1100$ or $\star 00 \star$. Therefore, we can say that if  $00$ is not present in the string 
$a_1a_2 \ldots a_n$, it will not appear in its consecutive images  under $\f$. Thus, 
block $1101$ will,  after each iteration of $f$, remain at the end, and will never be preceded by two zeros.
Eventually, after $n$ iterations, it will produce $101$, as shown in the example below.
\begin{displaymath}
\begin{array}{lllllllllllll}
1&0&1&0&0&1&1&1&0&\textbf{1}&\textbf{1}&\textbf{0}&\textbf{1}\\
  &1&1&0&0&1&1&0&\textbf{1}&\textbf{1}&\textbf{0}&\textbf{1}&  \\
  &  &0&0&0&1&0&\textbf{1}&\textbf{1}&\textbf{0}&\textbf{1}& & \\
  & & & 0&0&1&\textbf{1}&\textbf{1}&\textbf{0}&\textbf{1} & & &\\
  & & & &0&\textbf{1}&\textbf{1}&\textbf{0}&\textbf{1} & & & &\\
   & & & & &1&0&1 & & & & &
\end{array}
\end{displaymath}

What is left to show is that not having $00$ in $a_1a_2 \ldots a_n$ is necessary. This is a consequence of the fact
that  $\f(\star00\star)=00$, which means that if $00$ appears
in a string, then it stays in the same position after the rule 172 is applied. Indeed, if we had a pair of adjacent zeros in 
$a_1a_2 \ldots a_n$, it would stay in the same position when $\f$ is applied, and sooner or later
block $1101$, which is moving to the left, would come to the position immediately following this pair, and would be destroyed in the next 
iteration, thus never producing $101$. Such a process is illustrated below, where after three iterations the block $1101$ is destroyed
due to ``collision'' with $00$. $\Box$
\begin{displaymath}
\begin{array}{lllllllllllll}
1&0&1&0&0&0&1&1&0&\textbf{1}&\textbf{1}&\textbf{0}&\textbf{1}\\
& 1&1&0&0&0&1&0&\textbf{1}&\textbf{1}&\textbf{0}&\textbf{1}&\\
 && 0&0&0&0&1&\textbf{1}&\textbf{1}&\textbf{0}&\textbf{1}&&\\
  &&& 0&0&0&\textbf{1}&\textbf{1}&\textbf{0}&\textbf{1}&&&\\
   &&&& 0&0&1&0&1&&&&\\
   &&&&&  0&1&1 &&&&&
\end{array}
\end{displaymath}

 \begin{proposition}\label{struct111}
 Block $b$ belongs to $\f^{-n}(111)$ if and only if it has the structure 
\begin{equation}
 b= \underbrace{\star \star \ldots \star}_{n-2}  a_1 a_2\ldots a_{n+5},
\end{equation}
 where  $a_i \in \{0,1\}$ for $i=1,\ldots, n$ and the string $ a_1 a_2\ldots a_n$ 
satisfies the following three conditions:
\begin{itemize}
\item[(i)] $a_ia_{i+1} \neq 00$ for all $i=2\ldots n+4$;
\item[(ii)]  $a_{n+3}a_{n+4}a_{n+5} \neq 110$ and  $a_{n+2}a_{n+3}a_{n+4} \neq 110$;
\item[(iii)]  if $a_1a_2 \neq 00$, then $a_{n+1}a_{n+2}a_{n+3} \neq 110$.
\end{itemize}
\end{proposition}
\begin{figure}
  \begin{center}
    \includegraphics[scale=0.8]{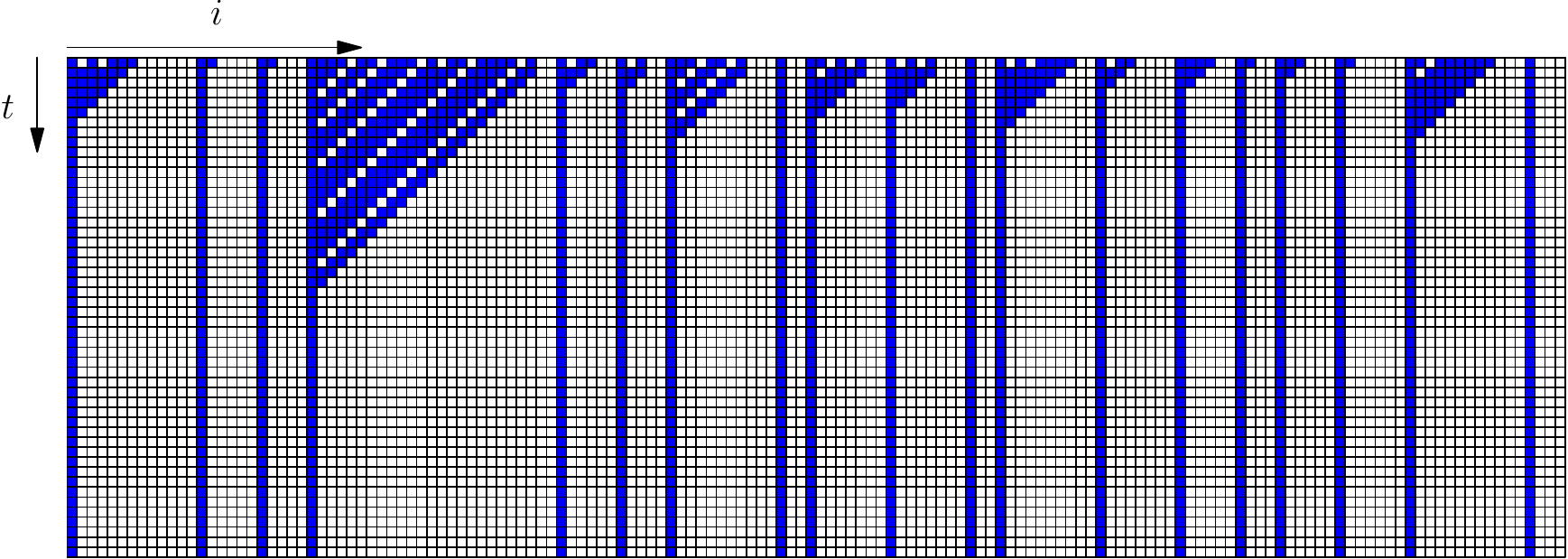}
  \end{center}
 \caption{Example of a spatiotemporal pattern produced by rule 172.}
  \label{figpattern}
\end{figure}
We will present only the main idea of the proof here, omitting some tedious details.
 It will be helpful to inspect spatiotemporal pattern generated by
rule 172 first, as shown in Figure~\ref{figpattern}. Careful inspection of this pattern
reveals three facts, each of them easily provable in a rigorous way:
\begin{itemize}
 \item[(F1)] A cluster of two or more zeros keeps its right boundary 
in the same place for ever.
 \item[(F2)] A cluster of two or more zeros extends its left boundary to the left one unit 
per time step as long as the left boundary is preceded by two
or more ones. If the left boundary it is preceded by 01, it stays in the same place.
 \item[(F3)] Isolated zero moves to the left one step at a time as long as it has at least two ones on the left.
If an isolated zero is preceded by 10, it disappears in the next time step.
\end{itemize}
Let us first prove that (i)-(iii) are necessary. Condition (i) is needed because if we had $00$ in the
string $a_2 \ldots a_{n+5}$, its left boundary would grow to the left and after $n$ iterations it would
reach sites in which we expect to find the resulting string $111$.

Moreover, string $a_1a_2 \ldots a_{n+5}$ cannot have $011$ at the end position, one site before the end, or two
sites before the end. If it had, 0 preceded by two 1's would move to the left   and, after $n$
iterations, it would reach sites where we want to find $111$. The only exception to this is
the case when $a_0a_1=00$. In this case, even if $011$ is in the second position from the end, it
will disappear in step $n-1$. This demonstrates that (ii) and (iii) are  necessary.

In order to prove sufficiency of (i)-(iii), let us suppose that the string $b$ satisfies all these conditions
yet $\f^n(b) \neq 111$. This would imply that at least one of the symbols of $\f^n(b)$ is equal to zero.
However, according to what we stated in F1--F3, zero can appear in a later configuration only as
a result of growth of an initial cluster of two of more zeros, or by moving to the left if it is
preceded by two ones. This, however, is impossible due to conditions  (i)-(iii). $\Box$.

\section{Enumeration of preimage strings}
Once we know the structure of preimage sets, we can enumerate them. For this, the following 
lemma will be useful.
\begin{lemma}\label{enumerationlemma}
The number of binary strings
$a_1a_2\ldots a_n$ such that $00$ does not appear as two consecutive terms $a_ia_{i+1}$ is equal to
$F_{n+2}$, where $F_n$ is the $n$-th Fibonacci number.
\end{lemma}
This result will be derived using classical transfer-matrix method. Let $g(n)$ be the number of binary strings
$a_1a_2\ldots a_n$ such that $00$ does not appear as two consecutive terms $a_ia_{i+1}$. We can think of such string 
as a walk of length $n$ on a graph with vertices $v_1=0$ and $v_2=1$ which has adjacency matrix $A$ given by
$A_{11}=0$,  $A_{12}=A_{21}=A_{22}=1$. One can prove that the generating function for $g$,
\begin{equation}
 G(\lambda)=\sum_{n=0}^\infty g(n+1)\lambda^n,
\end{equation}
can be expressed  by $G(\lambda)=G_{11}(\lambda)+G_{12}(\lambda)+G_{21}(\lambda)+G_{22}(\lambda)$, where
\begin{equation}
 G_{ij}=\frac{(-1)^{i+j} \det(I - \lambda A : j, i)}{\det(I-\lambda A)},
\end{equation}
and where $(M: j, i)$ denotes the matrix obtained by removing the $j-th$ row and $i-th$ column of $M$. Proof of this
statement can be found, for example, in~\cite{Stanley}.
Applying this to the problem at hand we obtain
\begin{equation}
G(\lambda)=\frac{-(2+\lambda)}{-1+\lambda+\lambda^2}.
\end{equation}
By decomposing the above generating function into simple fractions we get
\begin{equation}
G(\lambda)=\frac{\frac{3}{10} \sqrt{5}-\frac{1}{2}}{\lambda + \psi} + \frac{-\frac{1}{2}-\frac{3}{10} \sqrt{5}}{\lambda + 1- \psi},
\end{equation}
where $\psi = \frac{1}{2}+\frac{1}{2} \sqrt{5}$ is the golden ratio.
Now, by using the fact that 
\begin{equation}
 \frac{1}{\lambda+\psi}=-\sum_{n=0}^\infty \left( \frac{-1}{\psi}\right)^{n+1} \lambda^n,
\end{equation}
and by using a similar expression for $\frac{1}{\lambda+1-\psi}$, we obtain
\begin{equation}
G(\lambda)=\sum_{n=0}^\infty F_{n+3} \lambda^n,
\end{equation}
where $F_n$ is the $n$-th Fibonacci number, $\displaystyle F_n=\frac{\psi^n - (1-\psi)^n}{\sqrt{5}}$.
This implies that $g(n)=F_{n+2}$. $\Box$

\begin{proposition} \label{maincardin}
The cardinalities of preimage sets of $001$, $100$, $101$ and $111$ are given by
\begin{eqnarray}
  \card \f^{-n}(001) &=& 4^n,\\
  \card \f^{-n}(101) &=&2^{n-1}F_{n+2}, \\
 \card \f^{-n}(111) &=&2^n F_{n+3}. 
\end{eqnarray}
\end{proposition}
Proof of the first of these formulae is a straightforward consequence of Proposition~\ref{struct001}.
We have $2n$ arbitrary binary symbols in the string $b$, thus the number of such strings must be $2^{2n}=4^n$.

 The second formula can be immediately obtained using Lemma~\ref{enumerationlemma} and
Proposition~\ref{struct001}. Since
the first $n-1$ symbols of  $\f^{-n}(101)$ are arbitrary, and the remaining symbols form
a sequence of $n$ symbols without $00$, we obtain
\begin{equation}
 \card \f^{-n}(101) =2^{n-1} F_{n+2}.
\end{equation}

In order to prove the third formula, we will use Proposition~\ref{struct111}. We need to compute
the number of binary strings $a_1a_2\ldots a_{n+5}$ satisfying conditions (i)-(iii) of Proposition~\ref{struct111}.
Le us first introduce a symbol $\alpha_1 \alpha_2 \ldots \alpha_k$ to denote the string of length $k$
in which no pair $00$ appears. Then we define:
\begin{itemize}
 \item $A$ is the set of all strings having the form  $\alpha_1 \alpha_2 \ldots \alpha_{n+5}$,
\item $A_{1}$ is the set of all strings having the form $\alpha_1 \alpha_2 \ldots \alpha_{n+2}110$,
\item $A_{2}$ is the set of all strings having the form $\alpha_1 \alpha_2 \ldots \alpha_{n+1}1101$,
\item $A_{3}$ is the set of all strings having the form $\alpha_1 \alpha_2 \ldots \alpha_{n}11010$,
\item $A_{4}$ is the set of all strings having the form $\alpha_1 \alpha_2 \ldots \alpha_{n}11011$,
\item $B$ is the set of all strings having the form  $001\alpha_1 \alpha_2 \ldots \alpha_{n+2}$,
\item $B_1$ is the set of all strings having the form $001\alpha_1 \alpha_2 \ldots \alpha_{n-1}110$,
\item $B_2$ is the set of all strings having the form $001\alpha_1 \alpha_2 \ldots \alpha_{n-2}1101$.
\end{itemize}
The set $\Omega$ of binary strings $a_1a_2\ldots a_{n+5}$ satisfying conditions (i)-(iii) of Proposition~\ref{struct111}
can be now written as
\begin{equation} \label{bigset}
  \Omega= A\setminus(A_1 \cup A_2 \cup A_3 \cup A_4) \cup B\setminus(B_1 \cup B_2).
 \end{equation}
Since  $A_1\ldots A_4$ are mutually disjoint, and $B_1$ and $B_2$ are disjoint too, 
the number elements in the  set  $\Omega$ is 
\begin{align}
 \card \Omega = \card A - \card A_1 - \card A_2 - &\card A_3 - \card A_4 \\
&+ \card B - \card B_1 - \card B_2, \nonumber
\end{align}
which, using  Lemma~\ref{enumerationlemma}, yields
\begin{equation}
  \card \Omega =F_{n+7} -(F_{n+4}+F_{n+3}+F_{n+2} +F_{n+2}) + F_{n+4}-(F_{n+1}+F_n).
\end{equation}
Using basic properies of Fibonacci numbers, the above simplifies to $\card \Omega =4 F_{n+3}$.
Now, since in the Proposition~\ref{struct111} the string $a_1\ldots a_{n+5}$ is preceded by $n-2$
arbitrary symbols, we obtain
\begin{equation}
 \card \f^{-n}(111)=2^{n-2} \cdot 4F_{n+3}=2^n F_{n+3},
\end{equation}
what was to be shown.

\section{Density of ones}
\begin{figure}
  \begin{center}
    \includegraphics[scale=0.8]{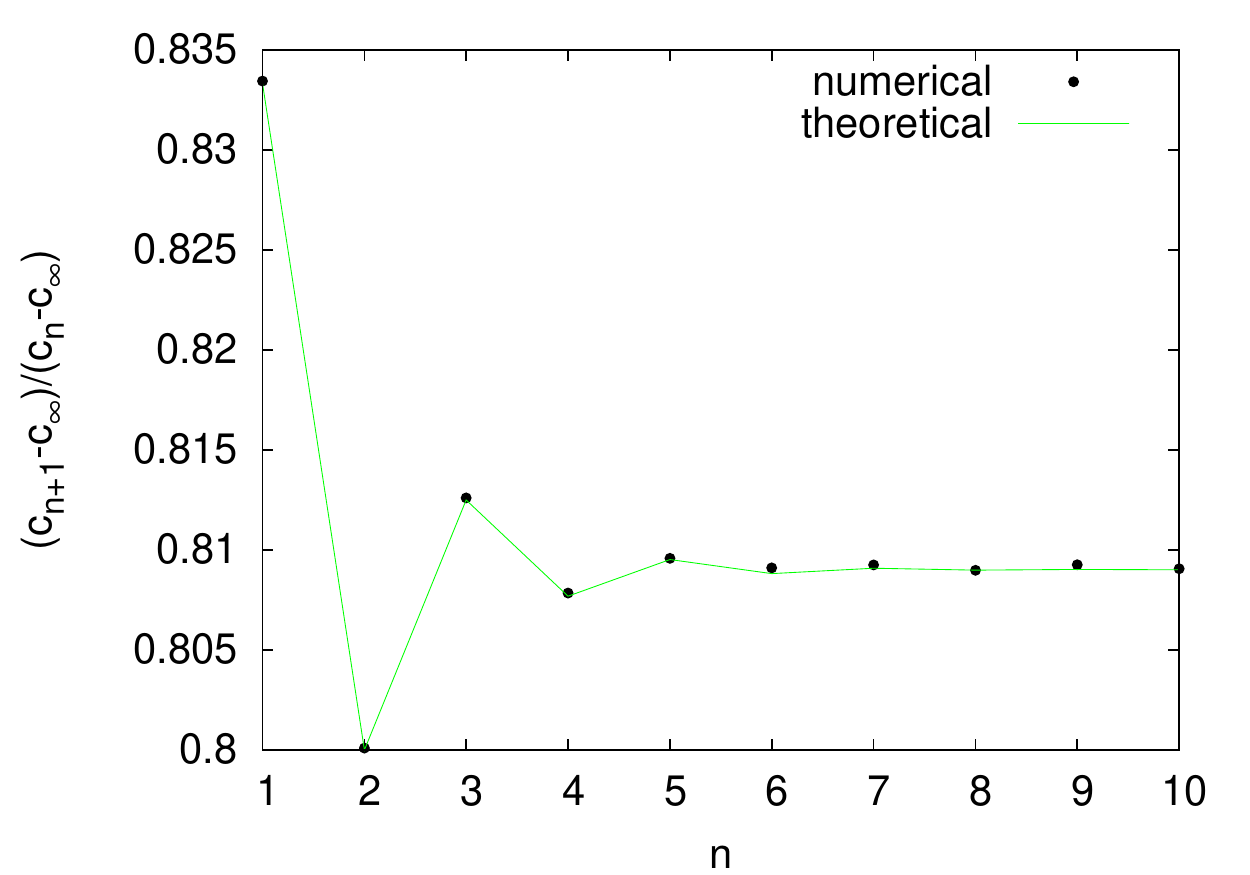}
  \end{center}
 \caption{Plot of the ratio $\displaystyle \frac{c_{n+1}-c_\infty}{c_{n}-c_\infty}$ as a function of time step $n$. Numerical
results were obtained by iterating rule 172 on a a configuration of length $10^8$ with periodic boundary
conditions.}
  \label{figdesnityplot}
\end{figure}
Using results of the previous section, eq. (\ref{densityformula}) can now be rewritten as 
\begin{equation}
 c_n=  \frac{4^{n-1} + 2^{n-1}F_{n+1}  +2^{n-1} F_{n+2}}{2^{2n+1}},
\end{equation}
which simplifies to
\begin{equation}
 c_n= \frac{1}{8}  + \frac{F_{n+3}}{2^{n+2}},
\end{equation}
or, more explicitly, to
\begin{equation}
c_n=\frac{1}{8} + \frac{(1+\sqrt{5})^{n+3} - (1-\sqrt{5})^{n+3}} {2^{2n+5} \sqrt{5}}.
\end{equation}
Obviously, $\lim_{n \rightarrow \infty} c_n=\frac{1}{8}$, in agreement with the numerical value reported
in \cite{Wolfram94}.
 We can see that $c_n$ converges toward
$c_\infty$ exponentially fast, with some damped oscillations superimposed over the exponential decay.
This is illustrated in Figure~\ref{figdesnityplot}, where, in order to emphasize the aforementioned oscillations,
instead of $c_n$ we plotted the ratio
\begin{equation} 
d_n= \frac{c_{n+1}-c_\infty}{c_{n}-c_\infty}
\end{equation}
as a function of  $n$. One can show that $d_n$ converges to the half of \textit{ratio divina} (golden ratio), $\psi/2 \approx0.809016\ldots$, as  illustrated in Figure~\ref{figdesnityplot}. We can see from this figure that the convergence is very fast and that the agreement between numerical simulations and the theoretical formula is nearly perfect. 

\section{Further results}
Results obtained in the previous two sections suffice to compute block probabilities for all blocks
of length up to 3. Proposition \ref{maincardin} together with eq. (\ref{blockprob}) yields formulas for
$P_n(001)$, $P_n(101)$, and $P_n(111)$. Consistency conditions give $P_n(01)=P_n(001)+P_n(101)$.
Furthermore $P_n(10)=P_n(01)$ due to the fact that $P_n(10)+P_n(00)=P_n(01)+P_n(00)=P_n(0)$.
Applying consistency conditions again we have $P_n(1)=P_n(10)+P_n(11)$, hence
$P_n(11)=P_n(1)-P_n(10)$, and, similarly, $P_n(00)=P_n(0)-P_n(10)$. This gives us probabilities of 
all blocks of length 2. Probabilities of blocks of length 3 can be obtained 
in a similar fashion:
\begin{align*}
 P_n(000)&=P_n(00)-P_n(100),\\
 P_n(110)&=P_n(11)-P_n(111),\\
 P_n(011)&=P_n(11)-P_n(111),\\
 P_n(010)&=P_n(01)-P_n(011).
\end{align*}
The only missing probability, $P_n(100)$ is the same as $P_n(001)$ because 
$P_n(100)+P_n(000)=P_n(001)+P_n(000)=P_n(00)$.
The following formulas  summarize these results. 
\begin{align} \nonumber
 P_n(000) &={5}/{8}-{2}^{-n-2}F_{ n+3} -{2}^{-n-4}F_{ n+2},  \\ \nonumber
P_n(001) &={1}/{8},\\ \nonumber
P_n(010) &={1}/{8}-{2}^{-n-3}F_{n+1 }, \\ \nonumber
P_n(011) &={2}^{-n-4}L_{ n+2 }, \\ \nonumber
P_n(100) &={1}/{8}, \\ \nonumber
P_n(101) &= {2}^{-n-4} F_{n+2 }, \\ \nonumber
P_n(110) &={2}^{-n-4}L_{ n+2}, \\ \nonumber
P_n(111) &={2}^{-n-3}F_{ n+3}, \nonumber
\end{align}
where $L_n=2F_{n+1}-F_n$ is the $n$-th Lucas number.
We can also rewrite these formulas in terms of cardinalities of preimage sets using
eq. (\ref{blockprob}), as  stated below.
\begin{theorem}\label{maintheorem}
Let $\f$ be the block evolution operator for CA rule 172.
Then for any positive integer $n$ we have
\begin{eqnarray*}
\card \f^{-n}(000)&=&5 \cdot 4^{n} - 2^{n+1}F_{n+3}-2^{n-1}F_{n+2},\\
\card \f^{-n}(001)&=&4^n,\\
\card \f^{-n}(010)&=&4^n-2^n F_{n+1},\\
\card \f^{-n}(011)&=&2^{n-1}L_{n+2},\\
\card \f^{-n}(100)&=&4^n,\\
\card \f^{-n}(101)&=&2^{n-1}F_{n+2},\\
\card \f^{-n}(110)&=&2^{n-1} L_{n+2},\\
\card \f^{-n}(111)&=&2^n F_{n+3},
\end{eqnarray*}
where $F_n$ is the $n$-th Fibonacci number, $\displaystyle F_n=\frac{\psi^n - (1-\psi)^n}{\sqrt{5}}$,  $\psi = \frac{1}{2}+\frac{1}{2} \sqrt{5}$, and $L_n$ is the $n$-th Lucas number,  $L_n=\psi^n +(1-\psi)^n$.
\end{theorem}

\section{Concluding remarks}
The method for computing block probabilities in cellular automata described in this paper
is certainly not applicable to arbitrary CA rule. It will work only if the structure of level sets
of preimage trees is sufficiently regular so that the level sets can be enumerated by 
some known combinatorial technique. Altough ``chaotic'' rules like rule 18, or complex 
rules such as rule 110 certainly do not belong to this category, in surprisingly many cases
significant regularities can be detected  in preimage trees.  Usually, this applies to
``simple'' rules, those which in Wolfram classification belong to class I, class II, and sometimes class III. Rule
172 reported here is one of the most interesting among such rules, primarily because the
density of ones does not converge exponentially to some fixed value as in many other cases,
but exhibits subtle damped oscillations on top of the exponential decay. Furthermore,
the appearance of Fibonacci and Lucas numbers in formulas for block probabilities
is rather surprising.

One should add at this point that the convergence toward the steady state can be slower than
exponential even in fairly ``simple'' cellular automata. Using similar method as 
in this paper, it has been found in  \cite{paper34} that in rule 14 the density of ones converges
toward its limit value approximately as a power law. The exact formula for the density of ones
in rule 14 involves Catalan numbers, and the structure of level sets is quite different than 
the one reported here.  Rule 142 exhibits somewhat similar behavior too, as reported in \cite{paper27}.

As a final remark, let us add that the results presented here assume initial measure $\mu_{1/2}$.
This can
be generalized to arbitrary $\mu_p$. In order to do this, one needs, instead of straightforward
counting of preimages, to perform direct computation of their probabilities using methods
based  on Markov chain theory. Work on this problem is ongoing and will be reported elsewhere.

\end{document}